# A Study on Features and Limitations of On-line C Compilers

Ramkumar Lakshminarayanan, PhD
Sur College of Applied Sciences, Oman.

Balaji Dhanasekaran, PhD
Salalah College of Applied Sciences, Oman

Ben George Ephrem, PhD
Higher College of Technology, Oman.

## ABSTRACT
Compilers are used to run programs that are written in a range of designs from text to executable formats. With the advent of the internet, studies related to the development of cloud based compilers are being carried out. There is a considerable increase of on-line compilers enabling on-line compilation of user programs without any mandate to. This study is specific to on-line C compilers to investigate the correctness, issues and limitations.

## 1. INTRODUCTION
Complete compiler infrastructure is just too complex to develop and maintain in the academic research environment [1]. When the field of compiling began in the late 1950's, its focus was limited to the translation of high-level language programs to machine code and to the optimization of space and time requirement. The most remarkable accomplishment, by far, of the compiler field is the widespread use of high-level languages. But over a period of time compilers are used for general purpose programming and mainly used for application development.

Compiler algorithms for parsing, type checking and inference, data flow analysis, loop transformations based on data-dependence analysis, register allocation based on graph coloring, and software pipelining are among the most elegant creations of computer science [1].

Compilers have been tested using randomized testing methods for nearly 50 years. A survey was done by Boujarwah and Saleh [8] and suggested methods for the compiler test case generation. Xuejun Yang et al created a randomized test-generator that supports compiler bug-hunting using differential testing [9]. Lindig [10] created a tool quest to create a randomly generated C program to find several compiler bugs. The quest has not used the control flow and arithmetic approach, creates complex data structures, loads them with complex data structures, loads them with constant values, and passes them to a function, where assertions check values received.

Improving the correctness of on-line C compilers is a worthy and important goal. C snippet is part of the trusted computing code for almost every modern computer system, including mission life critical pacemaker firmware. This tradition has extended its span and raids cloud based compilers that primarily deals with providing a platform to compile and execute programs that are independent of platform related restrictions. Cloud-based compilers can be used by any user who subscribes to it for a specific period of time. The functionalities that are provided by the cloud are to compile the programs on the go, file management and forums to discuss the issues [6].

The earlier testing studies focused on identifying the methods for testing the compilers using an automated approach as the control of compiler are in the control of the tester. The development of on-line compilers is heading towards the development of the cloud based compilers. Further to the need not to be installed, cloud compilers can easily be upgraded too.

## 2. C COMPILER
C is a general purpose programming language developed by Dennis Ritchie between 1969 and 1973. C is one of the programming the languages used over a period of time [2]. It is a compiler compatible to most architectures and operating systems. Many of the languages like C++, Java, JavaScript, C# and PHP have drawn many of the features from C. The current version of the standard is C11 approved on December 2011 [3]. The most common C Library is the C Standard Library, which is specified by the ISO and ANSI C Standards and comes with every C implementation [4].

## 3. ON-LINE COMPILERS (ONCOMP)
Table 1 lists mostly used on-line compilers and ordered [7].

**Table 1. List of on-line compilers**

| S.No | On-line Compiler |
|---|---|
| 1 | http://www.compileon-line.com/compile_c_on-line.php  OnComp 1 |
| 2 | http://code.hackerearth.com/  OnComp 2 |
| 3 | http://codepad.org/  OnComp 3 |
| 4 | https://ideone.com  OnComp 4 |
| 5 | http://www.on-linecompiler.net/  OnComp 5 |
| 6 | gcc.godbolt.org  OnComp 6 |
| 7 | http://cmpe150-1.cmpe.boun.edu.tr/on-lineCompiler/parts/  OnComp 7 |
| 8 | http://codebunk.com/  OnComp 8 |

| 9 | http://www.botskool.com/ OnComp 9 |
|---|---|
| 10 | http://rextester.com/ OnComp10 |
| 11 | Learn2Code (Chrome Plugin) OnComp 11 |

In the given listing OnComp 1 was designed to compile C and other programming languages. The version of the C compiler is gcc Version 4.8.1. Supports the ACE editor, VIM editor and Emacs editor. Has got the facility to compile multiple files. It is possible to download the source and the object files. The results of the application are displayed in the same screen. It provides the facility to give command line arguments and stdin input.

OnComp 2 is also designed to work with other programming languages. The application has not provided any details about the version of the compiler. In this application, it is possible to clone the code, share the code and download the source code. It supports API and VIM plugin. Login feature is included using the Facebook, Google and GitHub.

OnComp 3 works with the Sphere EngineTM. It provides the facility to track the code and possible to give the input and get the output in the same screen. It is possible to share and embed the source code. There are memory constraints of the submitted code and compilation time exceeds 10 seconds. Execution time, for unregistered users is 5 seconds, for registered, is 15 seconds. The size limit of program 64 kB. This compiler uses gcc version 4.8.1.

OnComp 4 does not have the feature to input the data and no information related to the compiler is provided. It is possible to create projects in the application.

OnComp 5 compiles the program on-line and the executable code is automatically downloaded to the user machine. It provides the compilation on Linux and Windows. The application is not provided with any details.

OnComp 6 compiles the program and provides the output in assembly code. It works with different compiler options like - 02. The compiler uses gcc 4.7.

The version of the OnComp 7 is not known and no details related to the web application are available. The code is compiled on-line and the executable is downloaded automatically.

OnComp8 supports 14 different languages other than C. It compiles and run code on-line. It is possible to chat,share code. It is possible to replay the history of the code. It is possible to create teams, create private code. This application was developed to support the interview process.

OnComp9' s compiler version is not provided. It is possible to provide input and the output is displayed in the same screen. The space provided for the output is very less. The application is loaded with a lot of advertisement a very poor design of user interface.

OnComp 10 is compiled using C version gcc 4.8.1 (gcc – Wall –std=gnu99 -02). The maximum allowed compile time is 30 seconds and after 10 seconds of execution, the process is killed. It provides API support and the possibility to create user.

OnComp 11 compiles and test the code within the chrome browser. The details related to the compiler are not provided and not available. It is also designed to support different programming languages. It reads input and writes output on the same screen.

Most of the compilers do not provide details of the operating system, architecture and version of the C compiler. The following code (Code 1) is used to identify the configuration of the system running on the on-line compilers [15].

**Code 1**

```
#include<sys/utsname.h>   /* Header for 'uname' */
main()
{
 struct utsname uname_pointer;
 uname(&uname_pointer);

 printf("System name - %s \n", uname_pointer.sysname);
 printf("Nodename    - %s \n", uname_pointer.nodename);
 printf("Release     - %s \n", uname_pointer.release);
 printf("Version     - %s \n", uname_pointer.version);
 printf("Machine     - %s \n", uname_pointer.machine);
}
```

Table 2 illustrates the system information of the on-line compilers that runs with linux operating system. Among the 11 on-line compilers, 4 of them use x86_64 server architecture, 2 uses i686 and others have not allowed the code to get compiled and not able to know the architecture and operating system

**Table 2. System Information Of The On-Line Compilers**

|   | On-line Compiler | System Information |
|---|---|---|
| 1 | http://www.compileon-line.com/compile_c_on-line.php OnComp 1 | System name - Linux Nodename - p3446206.pubip.serverbeach.com Release - 2.6.32-358.18.1.el6.x86_64 Version - #1 SMP Wed Aug 28 17:19:38 UTC 2013 Machine - x86_64 |
| 2 | http://code.hackerearth.com/ OnComp 2 | Not allowed the execution |
| 3 | http://codepad.org/ OnComp 3 | System name - Linux Nodename - 86828ea41af7 Release - 3.11.0-15-generic Version - #25-Ubuntu SMP Thu Jan 30 17:22:01 UTC 2014 Machine - x86_64 |
| 4 | https://ideone.com OnComp 4 | System name - Linux Nodename - checker Release - 2.6.34 Version - #6 SMP Fri Jan 21 15:21:52 CET 2011 Machine - i686 |
| 5 | http://www.on-linecompiler.net/ OnComp 5 | Not allowed the execution |
| 6 | gcc.godbolt.org OnComp 6 | Not allowed the execution |
| 7 | http://cmpe150-1.cmpe.boun.edu.tr/on-lineCompiler/parts/ | Not allowed the execution |
| | OnComp 7 | |
| 8 | http://codebunk.com/ <br> OnComp 8 | System name - Linux <br> Nodename - none <br> Release - 2.6.36.1 <br> Version - #1 Tue Dec 7 23:10:05 UTC 2010 <br> Machine - i686 |
| 9 | http://www.botskool.com/ <br> OnComp 9 | Not allowed the execution |
| 10 | http://rextester.com/ <br> OnComp 10 | System name - Linux <br> Nodename - vmi16328.contabo.net <br> Release - 3.11.0-17-generic <br> Version - #31-Ubuntu SMP Mon Feb 3 21:52:43 UTC 2014 <br> Machine - x86_64 |
| 11 | Learn2Code (Chrome Plugin) <br> OnComp 11 | System name - Linux <br> Nodename - ip-10-10-167-8.ec2.internal <br> Release - 3.11.0-19-generic <br> Version - #33-Ubuntu SMP Tue Mar 11 18:48:34 UTC 2014 <br> Machine - x86_64 |

Most of the compilers do not have interfaces in the same way. Designs are not easily understandable at the outset. They are not designed in any standard user interface.

The compiler objective is to provide the output in the executable format [14], OnComp 5 and OnComp 7 provides the output in the executable format, whereas other compilers do not have the facility of creating executable file. OnComp 6 provides the output in the assembly code, whereas other compilers do not have that facility. For further testing the usage of standard library codes OnComp 6 was not considered as it provides the assembly code.

Earlier there are studies which are performed on the compilers to find bugs by performing access summary testing on randomly generated C programs [5]. On-line compilers are available on the web without any proper information and without clarity about the standard architecture. So it becomes challenging for the end user to identify the truthfulness of on-line compilers. Bugs are out of reach for current and future automated program-verification tools because specifications that need to be checked were never written down in a precise way. The approach to verification is very impractical; however, other methods for improving compiler quality can succeed[6]. So the possibility of testing compilers, on-line, is by manually entering codes in the interface provided. The initial test was to ensure that on-line compilers are executing c programs.

Initially, all the online compilers are verified whether they are executing the C programs and noted they are compiling and providing results. It is being noted OnComp 6 provided the output in the assembly code.

Also, testing is furthered to investigate whether on-line compilers are executing the programs using a specific standard library like assert.h and ctype.h.

The assert.h header file of the C standard library provides a macro called assert which can be used to verify the assumptions made by the program and print a diagnostic message if this assumption is false. The ctype.h header file of the C standard library provides several functions useful for testing and mapping characters. All the functions accepts "int" as a parameter, whose value must be EOF or representable as an unsigned char.

Table 3 lists the details regarding the compilation of the asset.h and ctype.h code.

**TABLE 3. Asset.H And Ctype.H Execution Details**

| S.No | On-line Compiler | Asset.h | Ctype.h |
|---|---|---|---|
| 1 | OnComp 1 | Compiled | Compiled |
| 2 | OnComp 2 | Took more compilation time | Took more compilation time |
| 3 | OnComp 3 | Not able to give input | Compiled |
| 4 | OnComp 4 | Compiled | Compiled |
| 5 | OnComp 5 | Compiled | Compiled |
| 6 | OnComp 6 | Not tested | Assembly Code |
| 7 | OnComp 7 | Compiled | Compiled |
| 8 | OnComp 8 | No input option - but compiled without error | Yes |
| 9 | OnComp 9 | Compiled | Compiled |
| 10 | OnComp10 | Compiled | Compilation time exceeded 10 sec. |
| 11 | OnComp11 | Compiled | Compiled |

Among the 11 OnComp7 provided the expected output, one of the drawbacks in 2 of the on-line compilers is in providing the inputs. OnComp 2 took more compilation time and it varied from time and time. For ctype.h code, other than the OnComp 2 and OnComp 10, the code was compiled within the time and shown the expected result.

Sample code using float.h, limits.h, math.h, setjmp.h and signal.h were compiled using on-line compilers and the compilation details are provided in Table 4. The sample code containing float.h and limits.h was compiled with all the compilers. As in the earlier execution OnComp 2 took more time to compile. math.h code was not compiled by OnComp 8 and has not provided the expected result. setjmp.h code showed error in the OnComp 5 and OnComp 9. signal.h code was not compiled in OnComp 3, OnComp 9 and OnComp 10.

**TABLE 4. float.h, limits.h, math.h, setjmp.h And signal.h Execution Details**

| S.No | On-line Compiler | float.h | limits.h | math.h | setjmp.h | signal.h |
|---|---|---|---|---|---|---|
| 1 | OnComp 1 | Yes | Yes | Yes | Yes | Yes |
| 2 | OnComp 2 | More Time | More Time | Yes | Yes | Yes |
| 3 | OnComp3 | Yes | Yes | No | Yes | Disallowed |
| 4 | OnComp 4 | Yes | Yes | Yes | Yes | Yes |
| 5 | OnComp 5 | Yes | Yes | Yes | No | No |

| 6 | OnComp 6 | Assembly Code Output | | | | |
|---|---|---|---|---|---|---|
| 7 | OnComp 7 | Yes | Yes | Yes | Yes | Yes |
| 8 | OnComp 8 | Yes | Yes | No | Yes | Yes |
| 9 | OnComp 9 | Yes | Yes | Yes | No | No |
| 10 | OnComp 10 | Yes | Yes | Yes | Yes | No |
| 11 | OnComp 11 | Yes | Yes | Yes | Yes | Yes |

## 4. INDEFINITE LOOPS

One of the key challenges in the programs developed using the compilers is the handling of the indefinite loops and the common one is infinite looping. As a result, the program hangs.[16]. When the compiler is installed locally developer will have the command over killing the process of the program. By the following code (Code 2) we tested the on-line compiler's handling of indefinite loops.

**Code 2**

```
#include<stdio.h>
main()
{
while (1)
{
printf("Hello World");
} }
```

Table 5 provides the details of how the indefinite loops are handled by the on-line compilers.

**Table 5. Details Of Executing The Indefinite Loops**

| S.No | On-line Compiler | Details |
|---|---|---|
| 1 | OnComp 1 | Hanged / No Ouput |
| 2 | OnComp 2 | Hanged / No Output |
| 3 | OnComp 3 | Error. Time Out |
| 4 | OnComp 4 | Shown Output / Timeout |
| 5 | OnComp 5 | No issue for the on-line compiler as the executable is downloaded |
| 6 | OnComp 6 | Assembly Code Output |
| 7 | OnComp 7 | No issue for the on-line compiler as the executable is downloaded |
| 8 | OnComp 8 | Shown the output, truncated the output to 1000 characters. |
| 9 | OnComp 9 | Fatal error |
| 10 | OnComp 10 | Process killed after 10 seconds |
| 11 | OnComp 11 | Shown the output till the space is available. |

## 5. System() function

The C library function int system(const char * command) passes the command name or program name specified by command to the host environment to be executed by the command processor and returns after the command has been completed. The following code (Code 3) is used to test the system() function.

**Code 3**

```
#include<stdio.h>
int main()
{
    system("ls");
    return(0);
}
```

Table 6 gives the compilation details of the system() function of on-line compilers. OnComp 3 and OnComp 9 disallowed the system() call and the security can be enforced by allowing the system call for the registered users.

**Table 6. Details Of Executing The System() Function**

| S.No | On-line Compiler | Details |
|---|---|---|
| 1 | OnComp 1 | Compiled and shown the files |
| 2 | OnComp 2 | Not Compiled |
| 3 | OnComp 3 | Disallowed System call |
| 4 | OnComp 4 | Compiled and shown the files |
| 5 | OnComp 5 | No issue for the on-line compiler as the executable is downloaded |
| 6 | OnComp 6 | Assembly Code Output |
| 7 | OnComp 7 | Compiled. |
| 8 | OnComp 8 | Compiled. No Output |
| 9 | OnComp 9 | Restricted |
| 10 | OnComp 10 | Compiled and shown the files |
| 11 | OnComp 11 | No output. |

## 6. FILE MANAGEMENT

A file represents a sequence of bytes, does not matter if it is a text file or binary file. C programming language provides access to high level functions as well as low level (OS level) calls to handle file on your storage devices. The following codes (Code 4 & 5) are used for testing file management operation involving on-line compilers. Table 7 lists the execution details of the file management code.

**Code 4**

```
#include <stdio.h>

main()
{
  FILE *fp;
  char buff[255];

  fp = fopen("/tmp/test.txt", "r");
  fscanf(fp, "%s", buff);
  printf("1 : %s\n", buff );

  fgets(buff, 255, (FILE*)fp);
  printf("2: %s\n", buff );

  fgets(buff, 255, (FILE*)fp);
  printf("3: %s\n", buff );
  fclose(fp);

}
```

**Code 5**

```
#include <stdio.h>

main()
{
   FILE *fp;

   fp = fopen("/tmp/test.txt", "w+");
   fprintf(fp, "This is testing for fprintf...\n");
   fputs("This is testing for fputs...\n", fp);
   fclose(fp);
}
```

More than 50 percent of on-line compilers do not support file management code.

**Table 7. Details Of Executing The File Management Code**

| S.No | On-line Compiler | Details |
|---|---|---|
| 1 | OnComp 1 | Not Compiled |
| 2 | OnComp 2 | Not Compiled |
| 3 | OnComp 3 | Compiled |
| 4 | OnComp 4 | Runtime error |
| 5 | OnComp 5 | Compiled |
| 6 | OnComp 6 | - |
| 7 | OnComp 7 | Compiled |
| 8 | OnComp 8 | Runtime error |
| 9 | OnComp 9 | Restricted |
| 10 | OnComp 10 | Compiled |
| 11 | OnComp 11 | Not Compiled |

The figure 1 shows the details of the execution of indefinite loop code, system() function code and the file management code. OnComp 2 have not executed any of the test codes. Unexpected result happened during the compilation of file management code in OnComp 4 and OnComp 8. OnComp 5 and OnComp 10 compiled the system() function code and the file management code. None of the on-line compilers considered supports all the code execution.

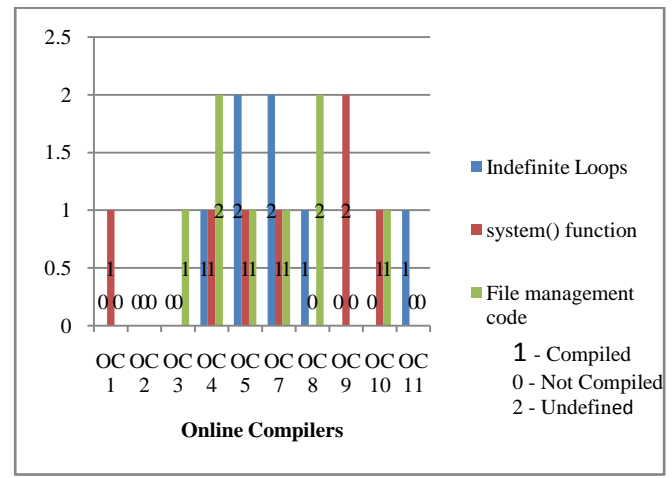

**Fig. 1 : Chart of execution details of indefinite loops, system() function and file management code.**

## 7. CONCLUSION

The basic idea of this study is to identify how on-line compilers handle the programs based on the C standard library, indefinite loops, system calls and file management. The initial challenge in the test of on-line compilers is to understand the user interface and on the understanding of the back-end compilers. One of the common issues noted in the study is the delay in the compile-time based on the internet bandwidth and some of the programs did not compile and there is no proper notification, even. Error rectification is possible only through compiler error notification that too is lacking in most of the compilers. The approach to developer notification must be improved by on-line compilers. Details like architecture, operating system and the compiler version are not mentioned either. Some on-line compilers have not compiled even the standard library codes. The indefinite loops were not managed well by most of the compilers. The system calls are supported in most of the compilers without registered users, which are likely to create security leaks. File management codes are not supported effectively in most of the compilers. There is a wider scope for fully designed on-line compilers. As a future study the existing C compiler has to be redesigned to support the cloud requirement and new testing approaches are to be designed

## 8. REFERENCES

[1] Mary Hall, David Padua and Keshav Pingali, "Compiler Research: The Next 50 years", Communications of the ACM, Vol. 52 : No 2, February 2009.

[2] http://www.tiobe.com/ accessed on December 2014.

[3] WG14 N1570 Committee Draft — April 12, 2011

[4] Stephen G, Programming in C (3rd Edition), July 18, 2004

[5] Eric Eide, John Regehr, Volatiles are Miscompiled, and what to do about it, Proceedings of the English ACM and IEEE International Conference on Embedded Software (EMSOFT), Atlanta, Georgia, USA, Oct 2008.

[6] Sajid Abdulla, Srinivasan Iyer, Sanjay Kutty, Cloud based compiler, International Journal of Students Research in Technology and Mangement, Vol(3), May 2013.